\documentclass[a4paper,prb,twocolumn,showpacs]{revtex4}
\usepackage{epsfig}
\usepackage{hyperref}
\begin{document}
\title{Electronic pseudogap of optimally doped
    Nd$_{2-x}$Ce$_x$CuO$_4$}
\author{D.~K.~Sunko}
\email{dks@phy.hr}
\author{S.~Bari\v si\'c}
\email{sbarisic@phy.hr}
\affiliation{Department of Physics, Faculty of Science, University of
Zagreb,\\ Bijeni\v cka cesta 32, HR-10000 Zagreb, Croatia.}
\newcommand{\ReS}{\mathrm{Re}\,\Sigma}
\newcommand{\ImS}{\mathrm{Im}\,\Sigma}
\newcommand{\eV}{\;\mathrm{eV}}
\begin{abstract}

We study the effect of antiferromagnetic correlations in the three-band Emery
model, in comparison with the experimental angle-resolved photoemission
(ARPES) spectra in optimally doped NCCO. The same calculation, formerly used
to describe BSCCO, is applied here, but in contrast to BSCCO, where quantum
paramagnon fluctuations are important, the characteristic energy of the
dispersive paramagnons in NCCO is of the order of $T_c$. The wide dispersing
features of the single-electron spectrum in NCCO are analogous to the BSCCO
hump. The Fermi surface is pseudogapped in both the nodal and antinodal
directions, although the detailed features differ, being dominated by loss of
intensity in the nodal direction, and loss of coherence in the antinodal one.
Direct oxygen-oxygen hopping is important in NCCO as well as in BSCCO, in
order to obtain overall agreement with the measured ARPES spectra.

\end{abstract}

\maketitle


High-temperature superconductivity (SC) is one of the premier problems of
physics today. It occurs across a broad range of copper oxide perovskites,
whose electronic responses are quite dissimilar. The $n$-doped
high-temperature superconductors show a number of outstanding differences with
the $p$-doped ones. First, $T_c$ is much lower, by a factor of 2--5. Also,
static antiferromagnetism (AF) extends to much higher doping, so that muon
resonance~\cite{Uefuji01} measurements clearly show a N\'eel temperature $T_N$
\emph{above} the superconducting (SC) $T_c$ in underdoped NCCO,
Nd$_{2-x}$Ce$_x$CuO$_4$ for $x<0.15$. At optimal doping, $x=0.15$, on which we
concentrate in the following, the static AF response abruptly disappears, but
another signal, indicating AF correlations without long-range order, persists
for $T<T_c$.\cite{Uefuji01,Yamada99} In contrast, the $p$-doped compounds have
a wide ``pseudogap'' region in the phase diagram, between the AF and SC
phases.\cite{Ding96,Loeser96}

The ARPES ``effective band'' electron response in the two classes of compounds
also appears to be quite different. In BSCCO and YBCO, the SC ARPES profile at
the van Hove (vH, antinodal) point is dominated by the so-called peak-dip-hump
structure.\cite{Campuzano99,Fedorov99,Lu01} There is a clear gap of the order
of several $T_c$, the ``leading-edge scale,'' which tapers to zero at the
nodal point in optimally doped (OP) BSCCO, but persists in the slightly
underdoped case, albeit without the narrow peak. In NCCO, a wide structure
disperses quickly across the Fermi level at both nodal and antinodal points. A
visible pseudogap persists only at the ``hot spot'' for AF scattering, where
the ARPES intensity drops sharply as the Fermi surface crosses the
skew-diagonal of the
zone.\cite{Armitage01,Onose01,Armitage01-1,Sato01,Armitage02,Blumberg02} This
is a precursor of the AF gap, which opens on the skew-diagonal for $T<T_N$.
The Fermi surface surmised from these measurements is hole-like in both BSCCO
and NCCO, but with a different shape, a rounded square around the $(\pi,\pi)$
point for BSCCO, and nearly a circle for NCCO.

These results are discussed here within the three band Hubbard model (Emery
model),\cite{Emery87,Varma87} which treats the $p$- and $n$-doped planar
CuO$_2$ cuprates on equal footing from the outset. It introduces the
difference $\Delta_{pd}$ between the O and Cu site energies, and the Cu--O and
O--O hoppings, $t_{pd}$ and $t_{pp}$ respectively, together with the large
interaction $U_d$ between the two holes on the Cu site. These parameters
reflect the chemistry of the CuO$_2$ planes and are therefore expected to be
similar for a given material upon both $p$ ($\delta>0$) and $n$ ($\delta<0$)
dopings. The physical regime which appears to be appropriate for high-T$_c$ SC
corresponds~\cite{Mrkonjic03,Mrkonjic03-2} to $\Delta_{pd}>|t_{pd}|>|t_{pp}|$,
with $|t_{pp}|$ and $t_{pd}^2/\Delta_{pd}$ of the same order of magnitude. The
AF order is suppressed asymetrically, typically for $\delta>0.01$ on the $p$
side and $-\delta>0.1$ on the $n$ side.

An alternative approach often applied to NCCO is the one-band Hubbard model,
viewed ``from the insulator.''\cite{Kuroki99,Kusko02,Kyung04} It attempts to
face experiment by taking into account the oxygen degree of freedom at least
through a large $t'\cos k_x\cos k_y$ term, a contribution of the same symmetry
as direct O--O hopping.\cite{Mrkonjic03} Models with more than three bands
are also sometimes considered,\cite{Bansil05} while on the $p$-doped side
three-band approaches have been used to discuss (stripe)
inhomogeneities.\cite{Lorenzana02,Abbamonte05}

In the Emery model, the asymmetry of AF and other physical properties between
the $p$- and $n$-dopings comes from the different role of $U_d$ in
the two cases. It is conveniently observed in the mean-field (MF) slave-boson
(SB) approximation for $U_d\to\infty$, which replaces the ``bare'' chemical
parameters $\Delta_{pd}$ and $t_{pd}$ with their renormalized values
$\Delta_{pf}$ and $t$, respectively, with $t_{pp}$ unchanged, in an effective
free three-band model, with a definite dependence of the renormalized
parameters on both the chemical ones and on $\delta$. Although crude, the MF,
paramagnetic, translationally invariant version of this theory, combined with
harmonic fluctuations of the SB field around the MF saddle point,
gives~\cite{Sunko05-1,Mrkonjic03} physically reasonable band structures in the
metallic regime.\cite{Lorenzana02}

The possible three-band regimes in the Emery model in such a situation have
been extensively classified.\cite{Mrkonjic03} In the presence of the
unrenormalized direct oxygen-oxygen hopping $t_{pp}$, the signature of strong
renormalization is a regime change,\cite{Mrkonjic04-1} from
$0<-t_{pp}<t_{pd},\Delta_{pd}/4$ to $-t_{pp}>t,\Delta_{pf}/4>0$. In
particular, $\Delta_{pf}< 4|t_{pp}|$ means that the intrinsic band-width
of the oxygen band exceeds the effective copper-oxygen splitting, so there
occurs an ``anticrossing,'' in which the lowest (bonding) hole band acquires a
significant oxygen character.

Although the main effects of $U_d$ on the overlap $t$ are similar on the $p$-
and $n$-sides, the renormalization mechanism is different. In the former,
paramagnetic charge interactions push the doped holes onto the oxygen sites.
This leads to a paramagnetic ``resonant band'' regime, $\Delta_{pf}\approx
4|t_{pp}|$, useful in BSCCO.~\cite{Sunko05-1} In NCCO, doping puts carriers on
the copper sites. Long-range AF survives to large $n$-dopings, but at some
point there are enough carriers for a significant gain in kinetic energy, if
they could spend more time on the oxygens. This is expressed by a band
renormalization into the anticrossing regime, $-4t_{pp}\gg\Delta_{pf}>0$,
which thus corresponds to a paramagnetic ``lower Hubbard band.''

The correlations omitted by the MF approach can also be described in terms of
the same bare parameters. In this scheme, far enough from the AF transition,
the magnetic correlations can be treated as dispersive paramagnons, and
included in a one-loop calculation:\cite{Sunko05-1}
$$
\Sigma_R(\mathbf{k},\omega)\propto-F
\int d^2q\int_{-\infty}^\infty d\omega'
$$
$$
\left[
\chi_R(\mathbf{Q+q},\omega-\omega')(1-f(\omega'))
\mathrm{Im}\,G^{(0)}_R(\mathbf{k-q-Q},\omega')
\right.
$$
\begin{equation}
\left. +G^{(0)}_R(\mathbf{k-q-Q},\omega-\omega')
n(\omega')\mathrm{Im}\,\chi_R(\mathbf{Q+q},\omega')\right],
\label{eqsigma}
\end{equation}
where $F$ is an effective interaction constant and $\mathbf{Q}=(\pi,\pi)$. The
first term is the magnon propagator convoluted with the electron response, the
second, vice versa. Both are equally important in BSCCO, but the one with the
boson occupation number dominates in NCCO.


\begin{figure}
\begin{tabular}{cc}
\hskip 3mm\raisebox{7mm}{\includegraphics[height=30mm]{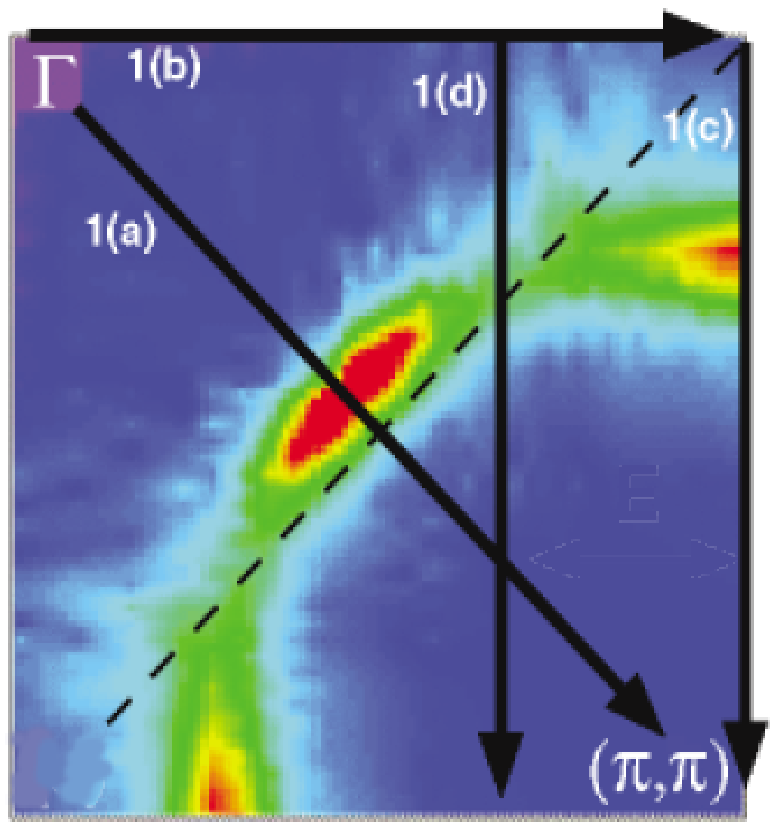}}&
\includegraphics[height=40mm]{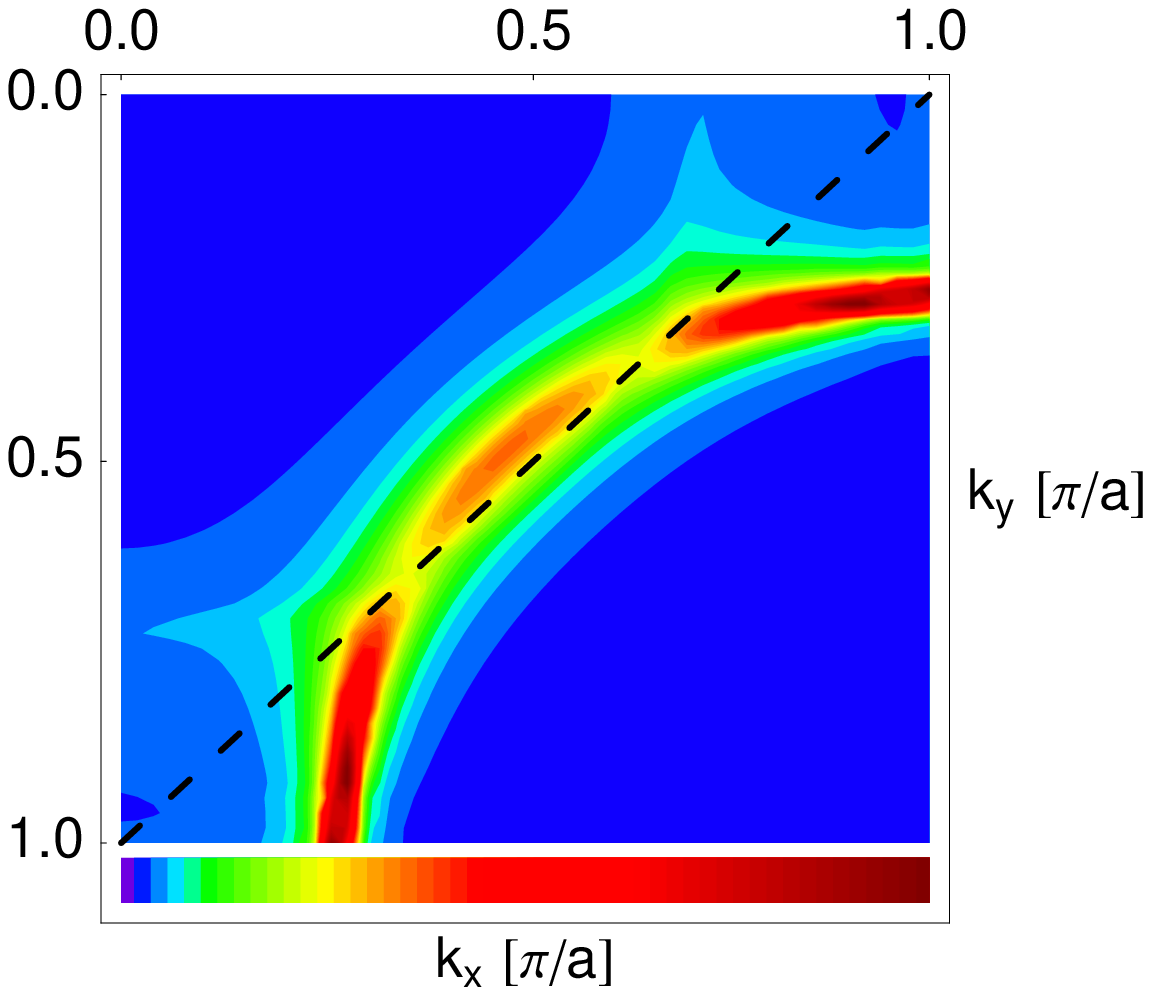}
\end{tabular}
\vskip -5mm
\caption{Left: experimental~\cite{Armitage01-1} NCCO ARPES intensity at the
Fermi level for $T<20$~K, $T_c=24$~K. Right: present calculation.}
\label{figfer}
\end{figure}
In Figure~\ref{figfer} we compare our results with experiment. The Fermi
surface is first fitted to obtain the renormalized parameters of the
three-band model. We can easily reproduce the shape of the Fermi surface in
NCCO in a regime very similar to the one in BSCCO,\cite{Sunko05-1} namely with
$\Delta_{pf}>-t_{pp}\gg t>0$, the main difference being that $\Delta_{pf}$ is
now smaller than in BSCCO: $\Delta_{pf}=1.6$~eV, $t_{pp}=-1.2$~eV and
$t=0.3$~eV, deep in the anticrossing regime. It is noteworthy that the
zeroth-order band prediction indicates that the effect of the oxygen hopping
$t_{pp}$ is as large in the $n$-doped cuprates as in the $p$-doped ones. The
Fermi surface in Fig.~\ref{figfer} is actually found beyond the anticrossing
point of the bonding dispersion, where the wave-functions have significant
oxygen character, making the argument in favor of anti-crossing
self-consistent. This regime cannot be reached consistently from the
single-band Hubbard model.

\begin{figure}
\begin{tabular}{cc}
\hskip -10mm\includegraphics[height=6.5cm]{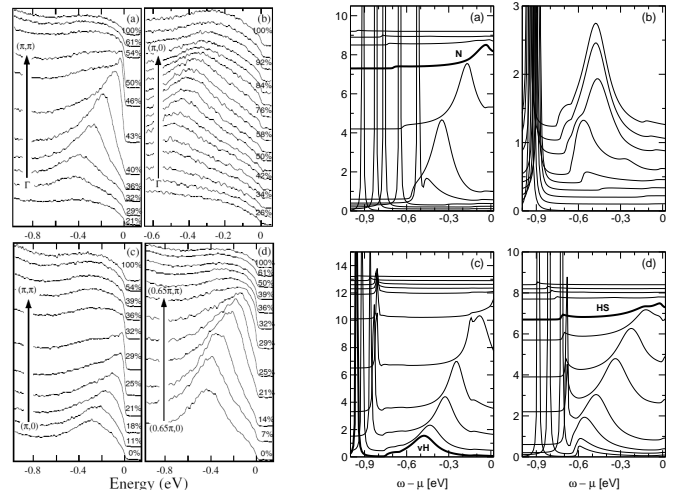}&
\hskip 0mm\includegraphics[height=6.5cm]{ncco-th}
\end{tabular}
\caption{Left: experimental~\cite{Armitage01-1} ARPES profiles for the cuts
(a)--(d) in Fig.~\ref{figfer}. The positions in the zone are indicated as
percentages along the ranges given by arrows. Right: calculated ARPES profiles
for the same positions, without the Fermi factor. The nodal $(\pi/2,\pi/2)$,
van Hove $(\pi,0)$, and hot-spot $(0.65\pi,0.36\pi)$ profiles are indicated
with heavy lines.}
\label{figedc}
\end{figure}

The paramagnon fit parameters are: band-edge
$\widetilde{\omega}=0.001\eV<kT=0.002\eV<\gamma=0.004\eV$ (damping), cutoff
$\omega_0=0.15\eV$, coupling constant $F=0.77\eV$, giving a self-energy range
shown in Fig.~\ref{sigma} below. The main input scales are all taken from
experiment.\cite{Uefuji01,Armitage01-1} The calculation is not sensitive to
the magnon damping and cutoff, kept roughly similar as in BSCCO. The only free
parameter left is the coupling constant $F$. With this one adjustment, all
ARPES scales, dispersions and widths in Fig.~2 were obtained rather well. Even
smaller experimental features have their counterparts in theory, such as the
flattening of the peak on the $(\pi,0)$--$(\pi,\pi)$ line, discussed below,
and the difference in dispersion shape (trend of the peaks) between panels (c)
and (d) in Fig.~\ref{figedc}.

The chemical potential (relative to the vH point) in this fit, $\mu=0.75$~eV,
is more than twice as large than experiment seems to suggest. The experimental
panel (c) in Fig.~\ref{figedc} shows that the ARPES profile at the vH point
has a maximum at about 0.35~eV binding energy. The lowest curve in the
theoretical panel (c) in the same figure explains the discrepancy: the
``bump'' there is in fact the upper wing of a spectral intensity split by AF
interactions. In this way the apparent Fermi level mismatch is resolved,
without spoiling the zeroth-order Fermi surface fit, as evident from
Fig.~\ref{figfer}. The narrow lower wing at nearly 1~eV binding corresponds to
the experimental ARPES intensity rising again at higher energy. In the
approach from the paramagnetic metal, high-energy features are subject to
further corrections, so let us first focus on the wide dispersive upper wings.

Our calculation in Fig.~\ref{figedc} gives the impression that all the wide
features observed in OP NCCO are incoherent.
Experimentally,~\cite{Armitage01-1} in addition to the hot spot, there is also
a drop in intensity in the nodal direction \emph{before} the Fermi level is
approached. This is exactly the opposite behavior than one would expect from a
quasiparticle. To emphasize the point, theoretical curves are shown without
the Fermi factor, and it is clear that the same phenomenon is occuring in the
calculation as well. We extend the experimental
observation,\cite{Armitage01-1} and claim that the Fermi surface in OP NCCO is
pseudogapped everywhere.

\begin{figure}
\includegraphics[height=50mm]{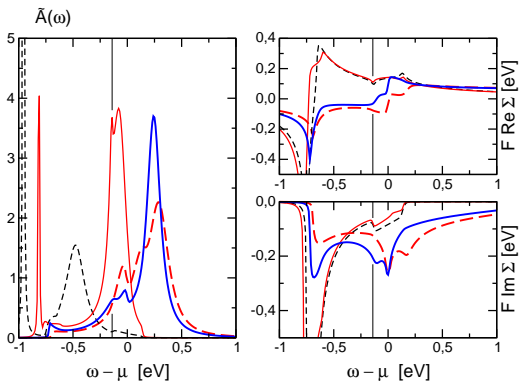}
\caption{Spectral intensities and the corresponding self-energies at selected
points in the Brillouin zone. Thin dashed line: vH point. Thin full line: 25\%
of the zone away from it in the direction of the AF point. Thick full line:
hot spot. Thick dashed line: 50\% of the zone in the nodal direction
$k_x=k_y$. The transition to the pseudogapped regime around $E_F$ is
emphasized by drop-down thin lines at $\omega-\mu=-0.14\eV$.}
\label{sigma}
\end{figure}
A detailed analysis is given in Figure~\ref{sigma}. At the vH point, the
chemical potential puts a quasiparticle at 0.75~eV binding energy, but the
self-energy shows that this is in the middle of a forbidden  region: $\ReS$ is
rising, $\ImS$ is large around there. The quasiparticle is split, and the wide
structure at 0.45~eV appears as the upper part of a split quasiparticle, not
necessarily incoherent. As $k_y$ increases along the $(\pi,0)$--$(\pi,\pi)$
line, this signal enters another forbidden region, at about $\pm 0.1$~eV
around $E_F$. We have purposefully put a rather low boson damping in the
calculation, to observe the moment of transition as a fine structure in the
profile at $k_y=0.25\pi/a$: the small sharp spike is at the edge of the
quasiparticle region, while the peak at right is already incoherent. (The
experimental profiles at 25\% and 29\% of the zone in panel (c) of
Fig.~\ref{figedc} show a similar flattening at the top.) Loss of coherence as
the Fermi energy is approached indicates a pseudogap regime, validating the
initial impression.

A similar phenomenon occurs in the nodal direction. The dominant feature is
loss of intensity rather than loss of coherence, because the Fermi surface is
nearly nested. The quasiparticle loses intensity before entering the pseudogap
region around $E_F$, and reappears at the other side, without really crossing
$E_F$. Succintly, the antinodal region is pseudogapped because an incoherent
signal crosses the Fermi level, while the nodal one is pseudogapped because a
coherent signal does not cross it. Technically, $\ReS$ increases everywhere at
the Fermi level, precluding a quasiparticle, while $\ImS$ is large at the
nodal Fermi crossing, because of nesting, but fairly small at the antinodal
crossing, which is not nested.

At the hot spot, both effects combine: an incoherent structure is suppressed
in intensity before it can cross the Fermi level. Finally, as one moves from
the vH point to the $\Gamma$-point, the increase in binding energy is
accompanied with a loss of both coherence and intensity, as the upper wing of
the split quasiparticle approaches the forbidden region below $\sim$0.5~eV
binding, where the slope $\ReS$ is positive, and $\ImS$ is large (see panel
(b) of Fig.~\ref{figedc}).


The present work establishes a number of parallels and distinctions between
BSCCO and NCCO. The chemical potential, relative to the vH point, and the
paramagnon coupling constant are much larger in NCCO, the first by a factor of
$\sim$30, the second $\sim$5. This makes the main ARPES scales in NCCO
significantly larger, $\sim$$0.5$ \emph{vs.} $\sim$$0.1$~eV in BSCCO. The
larger coupling constant is a plausible consequence of doping on the coppers,
where the carriers enter an otherwise strongly AF-polarized environment, as
evident from the wider extension of AF on the $n$-doped side. On the other
hand, when enough carriers are doped on the oxygens, the superexchange is only
indirectly involved in their magnetic correlations.

The paramagnon physical regime turns out to be significantly different in the
two pseudogaps. In SC BSCCO, paramagnons are quantum fluctuations, whose
energy scale (41~meV) is much higher than $T_c$. To put NCCO in the same
regime requires a paramagnon band-edge  $\widetilde{\omega}\sim 10$~meV. We
find this at odds with the distinct hot spot in experiment. If the band-edge
is comparable with the temperature, and especially if it is lower as here, the
hot spot becomes clearly discerned.\cite{Sunko05-2} We conclude the
paramagnons in SC NCCO are semiclassical, or transitional (the results are
similar if $\widetilde{\omega}\sim T_c$). Paramagnons in the vicinity of the
AF state indeed have low energies.~\cite{Uefuji01}

Conversely, if the paramagnon band-edge is lowered in BSCCO, the narrow
``antiadiabatic''~\cite{Sunko05-1} peak near the vH point disappears. This is
observed in underdoped BSCCO, as expected if the AF transition is approached.
We have argued~\cite{Sunko05-1} that the low-energy peak only obscures the
pseudogap in OP BSCCO, because it is part of the pseudogap profile, not a true
quasiparticle.  Without it, there remains the usual splitting of the spectral
intensity in two peaks, due to AF scattering. In BSCCO, the splitting
straddles the Fermi level, and the lower (occupied) wing is sufficiently close
to $E_F$ at the $(\pi,0)$--$(\pi,\pi)$ line to be incoherent. In NCCO, the
chemical potential places both wings below the Fermi energy at the vH point,
and the upper one becomes incoherent as it disperses towards $E_F$ along the
same line. Hence the wide dispersive wing in NCCO is analogous to the hump in
BSCCO.

Paramagnon damping was a critical parameter in BSCCO, where a switch from
over- to underdamped produced the change in ARPES profile observed at
T$_c$.\cite{Sunko05-1} By contrast, the parametrization used here for NCCO is
only weakly sensitive to the damping regime, which neatly parallels the
experimental situation: the changes in ARPES profiles as NCCO becomes
superconducting are very slight.\cite{Armitage01,Armitage01-1} This
strenghtens our assertion~\cite{Sunko05-1} that SC has also an indirect effect
on the ARPES spectra, by gapping out the magnon damping.

In both the BSCCO and NCCO calculations, the region of about $\pm 150$~meV
around $E_F$ does not support a quasiparticle, a scale set by the AF
interaction. Other sources of decoherence are absent in the numerics, coupling
to Mott charge fluctuations and to stripes.~\cite{Kivelson03} Both
are a natural next step in the present framework.\cite{Sunko05-1}
Thus Mott fluctuations are expected to account for the rise in response
below 0.5~eV binding, found in experiment, by broadening the narrow deeply
bound peaks visible in the calculation of Fig.~\ref{figedc}. A similar role of
Mott fluctuations at the $\Delta_{pf}$ scale has been found in the extended
Raman background, in the non-crossing approximation.\cite{Niksic95}

We expect our predictions to remain robust at less than 0.5~eV binding.
The lowest-energy feature is the coherent-incoherent crossover at
$\sim$0.15~eV, still high above the SC scale. Neither high-energy (Mott)
perturbations, operating at 1--2~eV, nor low-energy (SC) effects should
significantly change the picture given here in the energy window from T$_c$ to
$\sim$0.5~eV. The precise position of the crossover at 0.15~eV is
parameter-dependent, but its existence is not.

The intermediate-energy ($\sim$0.1 eV) features of ARPES spectra in both
BSCCO~\cite{Sunko05-1} and NCCO, down to the leading-edge scale at the
antinodal point in BSCCO, are reproduced by accounting for AF fluctuations
alone. If superconductivity, of whatever origin, is simply making use of the
available density of states, as calculated here, it obviously follows that
$T_c$ should be much higher in BSCCO than in NCCO. This indicates a somewhat
more complex structure of the SC gap than a single $d$-wave, not simply
related to the order parameter symmetry.

To conclude, we have accounted for ARPES spectra in NCCO focussing exclusively
on dispersive paramagnons, without low-energy magnetic responses, in the same
simple approach~\cite{Sunko05-1} as previously for BSCCO. We explain the
observed differences by a change in the physical regime of the paramagnons,
semiclassical in NCCO as opposed to quantum in BSCCO, and a much larger value
of the effective coupling constant. Both are consistent with the vicinity of
the AF transition in NCCO. Optimally doped NCCO is pseudogapped on the whole
Fermi surface, and the wide dispersive features are analogous to the hump in
BSCCO. Direct oxygen-oxygen hopping is as important in NCCO as in BSCCO. The
phenomenological value of the Cu--O hopping in NCCO indicates, as expected,
that it is renormalized by the strong on-site repulsion.


This work was supported by the Croatian Government under Project~$0119256$.

\end{document}